\documentclass[a4paper,11pt]{article}
\pdfoutput=1 

\usepackage{jheppub} 
\usepackage[utf8]{inputenc}
\usepackage[T1]{fontenc} 
\usepackage{booktabs} 
\usepackage{slashed}
\usepackage{placeins}
\usepackage{float}
\usepackage{changes}
\usepackage[]{units}
\usepackage{subfig}
\usepackage{pdflscape}
\usepackage{subfig}
\usepackage{marginnote}
\usepackage{tikz}
\usetikzlibrary{arrows}
\usetikzlibrary{shapes,arrows}
\usetikzlibrary{positioning}
\usetikzlibrary{fit}
\usetikzlibrary{shadows}

\usepackage{amsmath,amssymb,amsfonts}	
\usepackage{color}
\usepackage{amsmath}
\usepackage{amsfonts}
\usepackage{amssymb}
\usepackage{graphicx}
\usepackage{slashed}            
\usepackage{bbm}                
\usepackage{xspace}				
\usepackage{tikz}
\usetikzlibrary{arrows,shapes}
\usetikzlibrary{trees}
\usetikzlibrary{matrix,arrows} 				
\usetikzlibrary{positioning}				
\usetikzlibrary{calc,through}				
\usetikzlibrary{decorations.pathreplacing}  
\usepackage{pgffor}							
\usetikzlibrary{decorations.pathmorphing}	
\usetikzlibrary{decorations.markings}
\tikzset{
    vector/.style={decorate, decoration={snake}, draw},
	provector/.style={decorate, decoration={snake,amplitude=2.5pt}, draw},
	antivector/.style={decorate, decoration={snake,amplitude=-2.5pt}, draw},
    fermion/.style={draw=black, postaction={decorate},
        decoration={markings,mark=at position .55 with {\arrow[draw=black]{>}}}},   
    fermionbar/.style={draw=black, postaction={decorate},
        decoration={markings,mark=at position .55 with {\arrow[draw=black]{<}}}},
    fermionnoarrow/.style={draw=black},
    gluon/.style={decorate, draw=black,
        decoration={coil,aspect=0.5,amplitude=4pt, segment length=4pt}},
    scalar/.style={dashed,draw=black, postaction={decorate},
        decoration={markings,mark=at position .55 with {\arrow[draw=black]{>}}}},
    scalarbar/.style={dashed,draw=black, postaction={decorate},
        decoration={markings,mark=at position .55 with {\arrow[draw=black]{<}}}},
    scalarnoarrow/.style={dashed,draw=black},
    electron/.style={draw=black, postaction={decorate},
        decoration={markings,mark=at position .55 with {\arrow[draw=black]{>}}}},
	bigvector/.style={decorate, decoration={snake,amplitude=4pt}, draw},
	ghost/.style={dotted,draw=black, postaction={decorate},
        decoration={markings,mark=at position .55 with {\arrow[draw=black]{>}}}},
	ghostbar/.style={dotted,draw=black, postaction={decorate},
        decoration={markings,mark=at position .55 with {\arrow[draw=black]{<}}}},
         gluonloop/.style={decorate, draw=black,
        decoration={coil,aspect=0.6,amplitude=3pt, segment length=4pt}},   
}

\title{Confronting the coloured sector of the MRSSM with LHC data}

\author[a]{Philip Diessner,}
\author[b]{Jan Kalinowski,}
\author[c,1]{Wojciech Kotlarski\note{Corresponding author},}
\author[c]{and Dominik St\"ockinger}

\affiliation[a]{Deutsches Elektronen-Synchrotron DESY, Notkestraße 85, 22607 Hamburg, Germany\footnote{Former address}}
\affiliation[b]{Faculty of Physics, University of Warsaw, Pasteura 5, 02093 Warsaw, Poland}
\affiliation[c]{Institut f\"ur Kern- und Teilchenphysik, TU Dresden,
Zellescher Weg 19, 01069 Dresden, Germany}

\emailAdd{philip.diessner@desy.de}
\emailAdd{wojciech.kotlarski@tu-dresden.de}
\emailAdd{jan.kalinowski@fuw.edu.pl}
\emailAdd{dominik.stoeckinger@tu-dresden.de}
\preprint{arXiv:1907.11641}

\abstract{
R-symmetry leads to a distinct low energy realisation of SUSY with a significantly modified colour-charged sector featuring a Dirac gluino and scalar colour octets (sgluons).
In the present work we recast results from LHC BSM searches to discuss the impact of R-symmetry on the squark and gluino mass limits.
We work in the framework of the Minimal R-symmetric Supersymmetric Standard Model and take into account the NLO corrections to the squark production cross sections in the MRSSM that have become available recently.
We find substantially weaker limits on squark masses compared to the MSSM: for simple scenarios with heavy gluinos and degenerate squarks, the MRSSM mass limit is $m_{\tilde q} > 1.7$ TeV, approximately 600 GeV lower than in the MSSM.
}

\begin{document}
\maketitle
\flushbottom

\section{Introduction}

Supersymmetry (SUSY) is a new fundamental symmetry which extends Poincar\'e symmetry and links fermions and bosons.
It implies the absence of quadratic divergences to scalar masses, making the
electroweak scale technically natural.
SUSY remains one of the best motivated ideas for the physics beyond the Standard Model, even though no sign for SUSY has been found so far at the LHC.
The reason for the non-discovery of SUSY might be that SUSY is realized in a non-minimal form, which leads to weaker signals at the LHC and elsewhere as compared to the Minimal Supersymmetric Standard Model (MSSM).

The Minimal R-symmetric Supersymmetric Standard Model (MRSSM)~\cite{Kribs:2007ac} is such a well-motivated alternative to the MSSM.
It has more degrees of freedom but a higher degree of symmetry and therefore less free parameters.
It features a continuous unbroken R-symmetry~\cite{Fayet:1974pd} which does not commute with SUSY and, as an indirect consequence, Dirac instead of Majorana gauginos and $N=2$ SUSY multiplets in the gauge and Higgs sectors.

Thanks to the additional symmetry the phenomenology of the MRSSM is distinct from the one of the MSSM.
It constitutes a potential solution to the SUSY flavour problem; the R-symmetry effectively suppresses flavour-violating contributions in the quark/lepton sectors, thus allowing large flavour mixing and rather light masses in the squark/slepton sectors~\cite{Kribs:2007ac,Dudas:2013gga}.
In a series of papers \cite{Diessner:2014ksa,Diessner:2015yna,Diessner:2015iln,Diessner:2019ebm,Kotlarski:2019muo}, MRSSM electroweak phenomenology has been found to be rich and successful.
The model can accommodate the observed Higgs boson mass for stop masses in the $1$~TeV range, it is compatible with electroweak precision observables, and it can explain the observed dark matter relic density without conflict with direct dark matter searches and it can lead to interesting signals in muon g-2 and lepton flavour violation experiment.
Furthermore, refs.~\cite{Heikinheimo:2011fk,Herquet:2010ka,Kribs:2009zy,Chalons:2018gez,Kotlarski:2016zhv,Plehn:2008ae,Choi:2008ub,Kribs:2013oda,Kribs:2012gx} have investigated the coloured sector of the MRSSM or analogues models, based on tree-level calculations or estimates of NLO corrections; these references have shown that the mass limits in the MRSSM can be expected to be significantly weaker than in the MSSM.

In the present paper we focus on the coloured sector of the MRSSM in the context of LHC searches for new physics signals.
In particular we ask what are the LHC mass limits for squarks, assuming that squarks are the lightest coloured sparticles.
For this purpose we recast the LHC search results to the MRSSM.
Recently, a full NLO calculation of the squark production cross section in the MRSSM has become available \cite{Diessner:2017ske}.
The main differences between the MRSSM and MSSM predictions are as follows:\\
\phantom{M} (1) the R-symmetry forbids several important squark production
channels; the only allowed channels are $\tilde{q}_L\tilde{q}_R$ and $\tilde{q}_L\tilde{q}_L^\dagger$, $\tilde{q}_R\tilde{q}_R^\dagger$.
Because of this, the overall squark production rate is lower than in the MSSM.\\ \phantom{M}
(2) For the same reason, the overall K-factors in the MRSSM and MSSM are different; the difference typically amounts to $10$ to $20\%$. \\ \phantom{M}
(3) Specific properties of the MRSSM loop corrections, such as the Dirac nature of the gluino and the existence of the scalar gluon, can shift the K-factors up or down by ${\cal O}(10\%)$, depending on the involved mass ratios.

The present paper will provide updated limits on coloured sparticle masses, taking into account available LHC results and the NLO K-factors.
The outline of the paper is as follows.
In the next section we recapitulate the most important features of the strongly interacting sector of the MRSSM relevant to the present analysis and outline the computational framework.
Section 3 presents our results of recasting SUSY searches at the LHC in three distinct MRSSM squark scenarios.
Finally, conclusions are given in section 4.

\section{Relevant details of the MRSSM}
R-symmetry is an invariance under a continuous $U(1)$ transformation of Grassmannian coordinates $\theta\to e^{i\alpha}\theta$ (where, by convention, the R-charge of $\theta$ is assumed to be +1).
As a result, R-charges of component fields of a superfield differ by one unit.
Assigning charge $R=0$ to SM fields in the MRSSM implies non-vanishing R-charges for superpartners, therefore R-symmetry can be viewed as a continuous extension of the discrete R-parity.

One of the main motivations for R-symmetry is that it suppresses unwanted flavour- and CP-violating processes by eliminating trilinear soft SUSY-breaking scalar couplings.
In addition the symmetry forbids baryon- and lepton-number violating terms in the superpotential as well as dimension-five operators mediating proton decay.

Below we collect the relevant properties of the coloured sector of the MRSSM  and describe computational setup for recasting LHC search limits.

\subsection{Strongly interacting sector of the MRSSM}
R-symmetry implies in particular that Majorana gaugino masses are forbidden.
However, Dirac gaugino mass terms are allowed if additional adjoint chiral supermultiplets for each gauge group factor are introduced.
Concerning the strongly interacting sector of the MRSSM the list of states is as follows:
\begin{itemize}
\item
For each quark flavour $q_i$, there is a ``left-handed'' squark $\tilde{q}_{iL}$ with R-charge $R=+1$ and a ``right-handed'' squark $\tilde{q}_{iR}$ with R-charge $R=-1$.
Squarks of different type cannot mix, in contrast to the MSSM, but flavour mixing is possible.
The quark/squark supermultiplets are described by chiral superfields $\hat{Q}_{iL}$ and antichiral superfields $\hat{Q}_{iR}^\dagger$.
\item
The gluino is a Dirac fermion with R-charge $R=+1$.
Its left-handed component behaves like the MSSM gluino and is the
superpartner of the gluon, contained in a vector superfield $\hat{V}$.
The right-handed gluino component is part of an additional antichiral supermultiplet $\hat{O}^\dagger$; it has no direct
couplings to quarks or squarks.
\item
The model contains a scalar gluon (sgluon) field with R-charge $R = 0$ --- a complex scalar in the octet colour representation.
It is the scalar component of the $\hat{O}$ superfield.
The gluon, gluino and the sgluon together form an $N=2$ SUSY multiplet.
\end{itemize}

The SUSY Lagrangian and Feynman rules of this sector of the MRSSM can be found in ref.~\cite{Diessner:2017ske}.
Here we only provide the soft-breaking Lagrangian governing the mass terms
\begin{align}
\mathcal{L}_{\mathrm{soft}} = -\tfrac{1}{2} (m_{\tilde{q}_L}^2)_{ij}
\tilde{q}_{iL}^\dagger\tilde{q}_{jL} -
\tfrac{1}{2} (m_{\tilde{q}_R}^2)_{ij}
\tilde{q}_{iR}^\dagger\tilde{q}_{jR}
- m_{O}^2\left|O^{a}\right|^2 -
m_{\tilde{g}}\overline{\tilde{g}}\tilde{g} +m_{\tilde{g}}\left(\sqrt{2}D^a
O^a + \text{h.c.}\right),
\end{align}
where $i$ and $j$ are flavour indices running over the six flavours of up- and down-type quarks.
The squark mass matrices can have significant off-diagonal components in flavour space, leading to flavour-mixed squark mass eigenstates.
$D^a$ is the usual auxiliary field in the $SU(3)_C$ sector, and the complex sgluon field is decomposed as $O=({O_s + iO_p})/{\sqrt{2}}$, with the tree-level masses of components given by $m_{O_p}=m_{O}$ and $m_{O_s} = \sqrt{m_{O}^2 + 4 m_{\tilde{g}}^2}$.
The structure of the model, its mass matrices and $\beta$ functions motivate scenarios in which the squarks are the lightest coloured SUSY particles --- significantly lighter than gluinos and sgluons (see e.g. ref.~\cite{Diessner:2017ske} for detailed discussion).

\subsection{Tree-level squark production at the LHC}
\begin{figure}[!htp]
\begin{center}
\begin{tikzpicture}[line width=1.0 pt, scale=1.3, arrow/.style={thick,->,shorten >=2pt,shorten <=2pt,>=stealth}]
	\draw[fermion] (-1,0.5)--(-0.42,0);
	\draw[fermionbar] (-1,-0.5)--(-0.42,0);
	\node at (-1.2,0.5) {$u$};
	\node at (-1.2,-0.5) {$\overline{u}$};
	\draw[gluon] (-0.42,0)--(0.42,0);
	\draw[scalar] (0.42,0)--(1,0.5);
	\draw[scalarbar] (0.42,0)--(1,-0.5);
	\node at (1.5,0.5) {$\tilde{u}_L / \tilde{u}_R$};
	\node at (1.5,-0.5) {$\tilde{u}^\dagger_L / \tilde{u}^\dagger_R$};
\begin{scope}[shift={(4,0)}]
	\draw[gluon] (-1,0.5)--(-0.42,0);
	\draw[gluon] (-1,-0.5)--(-0.42,0);
	\node at (-1.2,0.5) {$g$};
	\node at (-1.2,-0.5) {$g$};
	\draw[gluon] (-0.42,0)--(0.42,0);
	\draw[scalar] (0.42,0)--(1,0.5);
	\draw[scalarbar] (0.42,0)--(1,-0.5);
	\node at (1.5,0.5) {$\tilde{u}_L / \tilde{u}_R$};
	\node at (1.5,-0.5) {$\tilde{u}^\dagger_L / \tilde{u}^\dagger_R$};
\end{scope}
\begin{scope}[shift={(8,0)}]
	\draw[fermion] (-1,0.5)--(0,0.5);
	\draw[fermion] (-1,-0.5)--(0,-0.5);
	\node at (-1.2,0.5) {$u$};
	\node at (-1.2,-0.5) {$u$};
	\draw[gluon] (0,0.5)--(0,-0.5);
	\draw[fermionnoarrow] (0,0.5)--(0,-0.5);
	\draw[scalar] (0,0.5)--(1,0.5);
	\draw[scalar] (0,-0.5)--(1,-0.5);;
	\node at (1.6,0.5) {$\tilde{u}_L/  \tilde{u}_R$};
	\node at (1.6,-0.5) {$\tilde{u}_R/ \tilde{u}_L$};
\end{scope}
\end{tikzpicture}
\end{center}
\caption{
Examples of tree-level diagrams for squark pair production in the MRSSM.
In contrast to the MSSM, R-symmetry forbids $\tilde{u}_L\tilde{u}_L$ or $\tilde{u}_L\tilde{u}_R^\dagger$ pairs to be produced.
For simplicity, only one (s)quark flavour is shown.
}\label{fig:TreeDiagrams}
\end{figure}
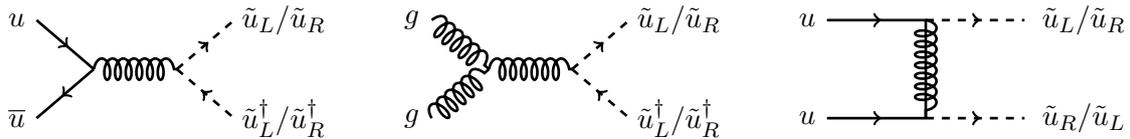

In the MRSSM all left/right-sfermions carry an R-charge of $\pm 1$, and in $pp$ collisions only final states with total $R=0$ are allowed.
As a result, in hard processes initiated by $q\bar{q}$ and $gg$ pairs produced squark and antisquark must have the same chirality while, in processes initiated by $qq$ pairs, produced squarks must have opposite chiralities as illustrated in figure~\ref{fig:TreeDiagrams}.
Since the number of possible squark-(anti)squark combinations is reduced, the squark production cross section in the MRSSM is expected to be smaller than in the MSSM for squarks of the same mass.
On the other hand, gluino production cross section is expected to be larger due to additional polarization states of Dirac gluinos.

\subsection{Calculational details and uncertainty estimation}

Our analysis amounts to a recasting of LHC search results into limits on squark and gluino masses both in the MSSM and the MRSSM.
It requires a precise evaluation of the relevant signal cross sections, a simulation of events, and the actual comparison to experimental data.
Much of the technical framework has been established in our previous works~\cite{Diessner:2014ksa,Diessner:2015yna,Diessner:2015iln,Diessner:2017ske}.
The MRSSM and MSSM mass spectrum generation is done using \texttt{SARAH-4.13.0} and \texttt{SPheno-4.0.3}~\cite{Staub:2008uz,Staub:2009bi,Staub:2010jh,Porod:2011nf,Staub:2012pb,Staub:2013tta,Goodsell:2014bna,Goodsell:2015ira}.
We use \texttt{Herwig-7.1.2}~\cite{Bellm:2015jjp,Bellm:2017bvx} for LO event generation of 13~TeV Supersymmetric-QCD (SQCD) events including subsequent decays.
Events for MRSSM and MSSM are generated using \texttt{UFO}~\cite{Degrande:2011ua} models generated by \texttt{SARAH}.
These events are then passed to \texttt{CheckMATE-2.0.26}~\cite{Drees:2013wra} to extract the limits for the considered parameter points.\footnote{With \texttt{CheckMATE} we make use of \texttt{Delphes}~3~\cite{deFavereau:2013fsa},
\texttt{FastJet}~\cite{Cacciari:2011ma},
the anti-$k_t$ clustering algorithm~\cite{Cacciari:2005hq,Cacciari:2008gp}.
}

Two important points are the inclusion of higher-order corrections and the estimation of remaining theoretical uncertainties.
We now describe our procedure for all involved processes, in turn for the MSSM and the MRSSM.

First we discuss the inclusion of higher-order corrections for the MSSM as it is straightforward with the use of the public software \texttt{NLLfast}~\cite{Beenakker:2015rna}, which includes the relevant NLO+NLL corrections.
Corrections of NNLL type are known for the MSSM and are included in the tool \texttt{NNLLfast} \cite{Beenakker:2016lwe}.
In line with the ATLAS analysis we make use of the global NLO K-factors provided by \texttt{NLLfast} for both squark as well as gluino final states to scale the prediction for the total production cross section appropriately as calculated by \texttt{Herwig} from LO to NLO.
We do not take the NLL K-factors into account for the direct exclusion limit in the MSSM as they are not available for the MRSSM and for the sake of comparison of both models we consider the same theoretical accuracy.
So far, no experimental analysis makes use of the NNLL K-factors.
Therefore, we only include them as part of our analysis of the theory uncertainty.

We arrive at an estimate of the theory uncertainty of the cross section prediction by noting that the NNLL corrections in the relevant regions are always substantially larger than the NLL corrections.
Furthermore, the NNLL corrections are larger than the uncertainties from scale and PDF variations.
Therefore, the non-inclusion of NNLL contributions is always the largest source of theoretical uncertainty dominating other effects.
For the uncertainty on the total cross section in the MSSM $\Delta \sigma^{\text{MSSM}}$ we take the absolute difference of the NLO to NNLL cross sections provided by \texttt{NNLLfast} summed over all processes:
\begin{equation}
\Delta \sigma^{\text{MSSM}} \equiv \left|\sigma_{\text{tot}}(\text{NNLL})-\sigma_{\text{tot}}(\text{NLO})\right|\;.
\end{equation}

The method used for the inclusion of higher-order corrections and the estimation of remaining theoretical uncertainties needs to be adapted when going to the MRSSM as several relevant effects are not as well known.
The squark/antisquark production processes in the MRSSM have been computed at NLO in ref.~\cite{Diessner:2017ske} using several methods.
We take into account these NLO corrections via global K-factors, evaluated numerically using \texttt{MadGraph5\_aMC@NLO-2.5.5}~\cite{Alwall:2011uj,Alwall:2014hca}%
\footnote{With \texttt{MadGraph5\_aMC@NLO} we make use of \texttt{MadFKS}~\cite{Frederix:2009yq},
\texttt{LHAPDF6}~\cite{Buckley:2014ana},
\texttt{MMHT2014} PDFs \cite{Harland-Lang:2014zoa}.
}
and \texttt{GoSam-2}~\cite{Cullen:2011ac,Cullen:2014yla}.%
\footnote{With \texttt{GoSam} we make use of the programs
\texttt{Qgraf}~\cite{Nogueira:1991ex},
\texttt{Form}~\cite{Vermaseren:2000nd},
\texttt{Ninja}~\cite{Mastrolia:2012bu,Peraro:2014cba,vanDeurzen:2013saa} (which
uses \texttt{OneLOop}~\cite{vanHameren:2010cp}) and
\texttt{Golem95}~\cite{Binoth:2008uq,Cullen:2011kv,Guillet:2013msa}.
We rely on the \texttt{BLHA2} interface~\cite{Binoth:2010xt,Alioli:2013nda}
as connection to \texttt{MadGraph5\_aMC@NLO}.
}

The theory uncertainty of the squark/antisquark production cross sections is dominated by two sources: (1) the unknown masses of sgluons, which impact the size of loop corrections, (2) the unknown NLL and NNLL corrections.

To estimate the theory uncertainty of type (1), we note that the influence of the sgluon mass on the MRSSM cross sections due to higher order effects has been studied in ref.~\cite{Diessner:2017ske}.
It was found there that a non-decoupling effect exists due the sgluon-gluino mass splitting.
We take this effect into account in our uncertainty by assuming a central octet mass of $m_O=10$~TeV and varying it by a factor of ten for an estimate on the sgluon mass dependence of the squark/antisquark production cross sections.

To estimate uncertainty of type (2) we need to estimate unknown higher order corrections in the MRSSM.
In the MSSM, it is known that the missing NNLL corrections are large --- significantly larger than the NLL ones.
As these corrections originate largely from QCD threshold production effects we assume that their effects in the MRSSM are of a similar magnitude and we may use the NLO to NNLL K-factor of \texttt{NNLLfast} as an approximation for the uncertainty sources (2).
This discussion can be summarized in the following formula for our uncertainty estimate

\begin{align}
\Delta \sigma^{\text{MRSSM}} =& \left(
\left(\frac{\sigma_{\text{tot}}^{\text{MSSM}}(\text{NNLL})-\sigma_{\text{tot}}^{\text{MSSM}}(\text{NLO})}{\sigma_{\text{tot}}^{\text{MSSM}}(\text{NLO})}\sigma_{\text{tot}}^{\text{MRSSM}}(\text{NLO})
\right)^2\right.\notag\\
&+\left.\left(\sigma_{\tilde q\tilde q+\tilde q\tilde q^*}(\text{NLO},m_O=m_O^{\text{up/down}})-\sigma_{\tilde q\tilde q+\tilde q\tilde q^*}(\text{NLO},m_O=m^{\text{ref}}_O)\right)^2 \right)^{1/2}\;.
\label{MRSSMuncert}
\end{align}
In the last line the reference octet soft breaking mass parameter is set to $m_O^{\text{ref}}=10$~TeV and $m_O^{\text{up/down}}$ corresponds to either 1 or 100~TeV (depending on which one gives the maximum contribution to the uncertainty)

Gluino production is of minor importance for the analysis of the MRSSM mass limits, since gluinos are likely to be significantly heavier than squarks~\cite{Kribs:2007ac} and no results for higher order corrections are available.
Hence, we approximate the MRSSM gluino production cross section K-factors by the MSSM ones for the analogues mass spectra.
The MSSM K-factors are again evaluated using \texttt{NLLfast}.
The theory uncertainty in the MRSSM and the MSSM is again approximated by the numerical value of the MSSM NNLL corrections evaluated using \texttt{NNLLfast}.
The effect of the sgluon mass as source of uncertainty is not taken into account.

A special scenario we study in this work is the squark/antisquark production for the special cases of non-degenerate squark masses and mixing between 1st and 3rd generation squarks.
As the higher order results of \texttt{NLLfast} and the work in ref.~\cite{Diessner:2017ske} are available for squarks with degenerated masses and zero mixing the relevant K-factors can only be approximated.
For the case of non-degenerated masses and zero mixing we use the global K-factors provided by \texttt{NLLfast} for the MSSM as well as our work for the MRSSM and use the smaller of the two squark masses
considered as squark mass input.
The same is done for estimation of the uncertainty.
When additionally the mixing between the first and third generation via an angle $\theta$ is allowed (assuming that quark masses can be neglected against the degenerate soft-breaking squark mass of one generation) the inclusion of higher order corrections requires further consideration.
The gluon-initiated production of squarks is flavour diagonal and the dependence on $\theta$ drops out when all masses are equal.
Then only the quark-initiated cross section shows a dependency of $\sin^4\theta$.
We assume that K-factors for the first generation and third generation are of similar size, not too far from one and multiply the total leading order cross section by a factor of \[\sin^4\theta\cdot K_{\text{first gen.}}+ (1-\sin^4\theta)\cdot K_{\text{third gen.}}\,.\]
The case of non-degenerate squark masses is handled the same way as the one without mixing.
We assume that only the lowest considered squark mass is relevant and take it as the one mass parameter needed for the available tools.

\texttt{CheckMATE} includes ATLAS analyses, which are sensitive to the final states of these processes.
The relevant ones, on which we will focus in the following section, are the two to six jets plus missing transverse energy search~\cite{Aaboud:2016zdn} and the search for stops~\cite{ATLAS-CONF-2017-019,Aaboud:2017ayj}.\footnote{\texttt{CheckMATE} includes additional experimental analyses and we have checked that the ones we list are the most sensitive.}

\section{Squark mass limits in the MRSSM}

In this section we present the limits on squark masses in the MRSSM and compare the exclusions to the case of the MSSM.
We distinguish three scenarios:
\begin{itemize}
\item no flavour mixing, L/R states mass degenerate
\item no flavour mixing, independent masses of L/R states
\item stop--squark mixing, equal in L/R sectors
\end{itemize}
Unless specified otherwise, the squarks are assumed to decay into a quark and the LSP, which is assumed to be massless.
This will provide the most stringent limit on the parameter space.

\subsection{
Scenario A: squarks of the first and second
  generation degenerate in mass, no flavour mixing}
\begin{figure}
\centering
\includegraphics[width=0.525\textwidth]{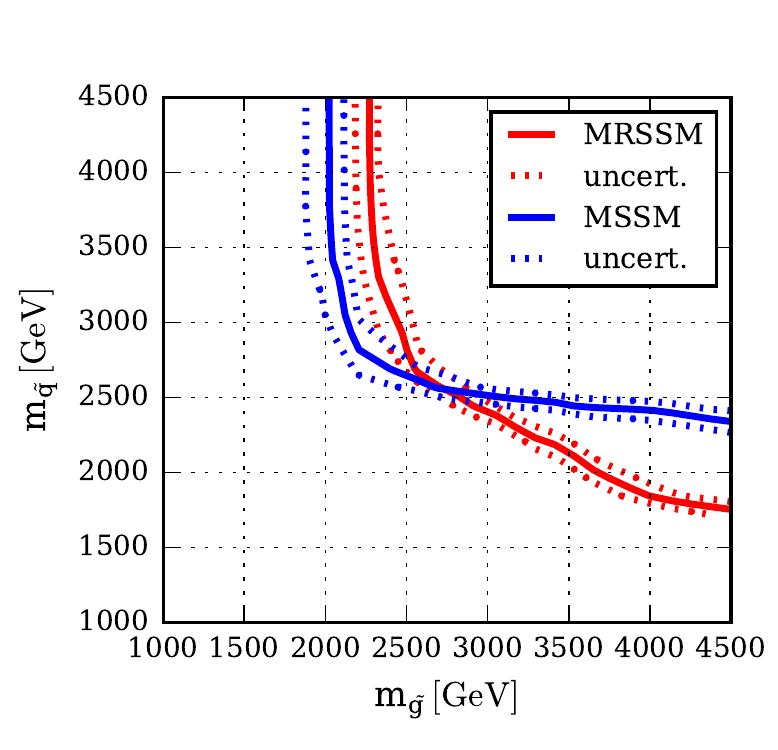}
\caption{
Mass limits on squarks and gluinos in case of degenerate 1st and 2nd generation squarks.
The limits are based on a recasting of LHC analyses, and the marked bands correspond to the theory uncertainty of the cross sections (see discussion in Sec. 2.3).
}
\label{fig:scenarioA}
\end{figure}

We begin with a simple but still particularly interesting scenario.
All masses of the first and second generation squarks are set to a common value of $m_{\tilde{q}}$, flavour mixing is zero and the LSP is massless.
Figure\ \ref{fig:scenarioA} shows the exclusion contours in the squark--gluino mass plane for the MRSSM and the MSSM.

We first focus on the region with very heavy gluino, $m_{\tilde{g}}>3.5$~TeV, where gluino final states are irrelevant.
This parameter region is particularly motivated in the MRSSM because supersoft supersymmetry breaking naturally allows for a small hierarchy between gaugino and scalar masses.
In this parameter region the most important production processes are the pure squark final states $\tilde{q}\tilde{q}$.
Production of $\tilde{q}\tilde{q}^\dagger$ and $\tilde{q}^\dagger\tilde{q}^\dagger$ pairs is less relevant because antisquark production requires antiquarks or gluons in the initial
state and is therefore suppressed by antiquark PDFs or, in the case of heavy squarks, gluon PDFs.

In the MRSSM only the $\tilde{q}_L\tilde{q}_R$ is allowed and the production process is therefore suppressed by two powers of $m_{\tilde{g}}$ due to the chiral structure of the quark-squark-gluino coupling and Dirac nature of the gluino.
In the MSSM, also $\tilde{q}_L\tilde{q}_L$ and $\tilde{q}_R\tilde{q}_R$ production is allowed and the gluino mass suppression is
weaker.
Accordingly, figure~\ref{fig:scenarioA} shows that the resulting MRSSM squark mass limits are substantially weaker.
For a reference gluino mass of 5~TeV, we obtain the following limits
\begin{align}
\label{eq:first_second_gen}
  m_{\tilde{q}} &> \left\{\begin{array}{l}
  1.7\text{ TeV (MRSSM)}\\
  2.3\text{ TeV (MSSM)}
  \end{array}\right.(m_{\tilde{g}}=5\text{ TeV})
\end{align}
This is one of the central results of the present paper.
The reason for the 600~GeV difference in limit is the reduced squark production cross section in the MRSSM compared to the MSSM.

For lighter gluino masses, further production processes of $\tilde{g}\bar{\tilde{g}}$ and $\tilde{q}\tilde{g}$ become relevant.
In particular pure gluino production is enhanced in the MRSSM compared to the MSSM, simply because the gluino has 4 instead of 2 degrees of freedom.
For this reason, the LHC reach for light gluinos is higher in the MRSSM.
The appropriate limits are
\begin{align}
  m_{\tilde{g}} &> \left\{\begin{array}{l}
  2.2\text{ TeV (MRSSM)}\\
  2.0\text{ TeV (MSSM)}
  \end{array}\right.(m_{\tilde{q}}=5\text{ TeV})
\end{align}

\subsection{Scenario B: mass splitting between left- and right-handed
  squarks, no flavour mixing}

\begin{figure}
\includegraphics[width=0.6\textwidth]{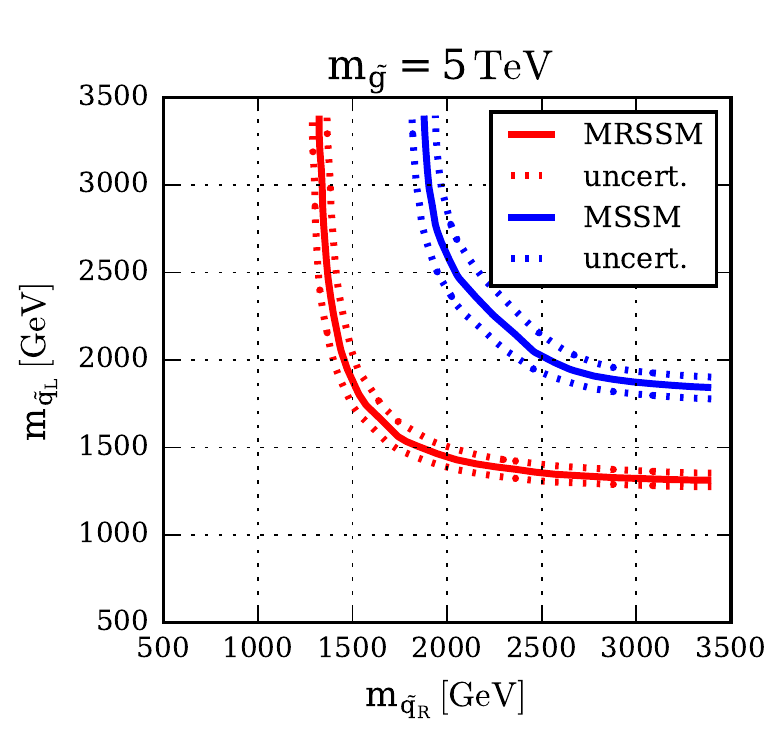}
\centering
\caption{
Mass limits on left-/right-squark masses (assuming universality among generations) for $m_{\tilde g} = 5$ TeV.
}
\label{fig:scenarioB}
\end{figure}
Now we assume one common left-handed squark mass $m_{\tilde{q}_L}$ and one common right-handed squark mass $m_{\tilde{q}_R}$ for the first and second generation squarks.
The gluino is assumed to be heavy ($m_{\tilde{g}}=5$ TeV) and the LSP to be massless. Figure\ \ref{fig:scenarioB} shows the resulting exclusion contour in the $\tilde{q}_L$--$\tilde{q}_R$ mass plane.
The exclusion is symmetric under the exchange $m_{\tilde{q}_L}\leftrightarrow m_{\tilde{q}_R}$; hence we focus our discussion only on the case $m_{\tilde{q}_L}\ll m_{\tilde{q}_R}$. In this case the most important final states are $\tilde{q}_L\tilde{q}_L$ and $\tilde{q}_L\tilde{q}_L^\dagger$.
In the MRSSM, only the second one is allowed but it is PDF-suppressed; hence the limit on the lighter squark mass is rather weak.
In the MSSM both final states are allowed, and the mass limit is much stronger. As a reference, for gluino and right-squark masses of 5~TeV, we obtain as limits
\begin{align}
  m_{\tilde{q}_L} &> \left\{\begin{array}{l}
  1.3\text{ TeV (MRSSM)}\\
  1.8\text{ TeV (MSSM)}
  \end{array}\right.(m_{\tilde{q}_R}=m_{\tilde{g}}=5\text{ TeV})
\label{eq:first_second_gen2}
\end{align}
Similar to eq. 3.1 the squark mass limits are about 500~GeV lower in the MRSSM.

\subsection{Scenario C: flavour mixing between first and third generation}
\begin{figure}
\centering
\subfloat[\label{fig:4a}]{\includegraphics[width=0.49\textwidth]{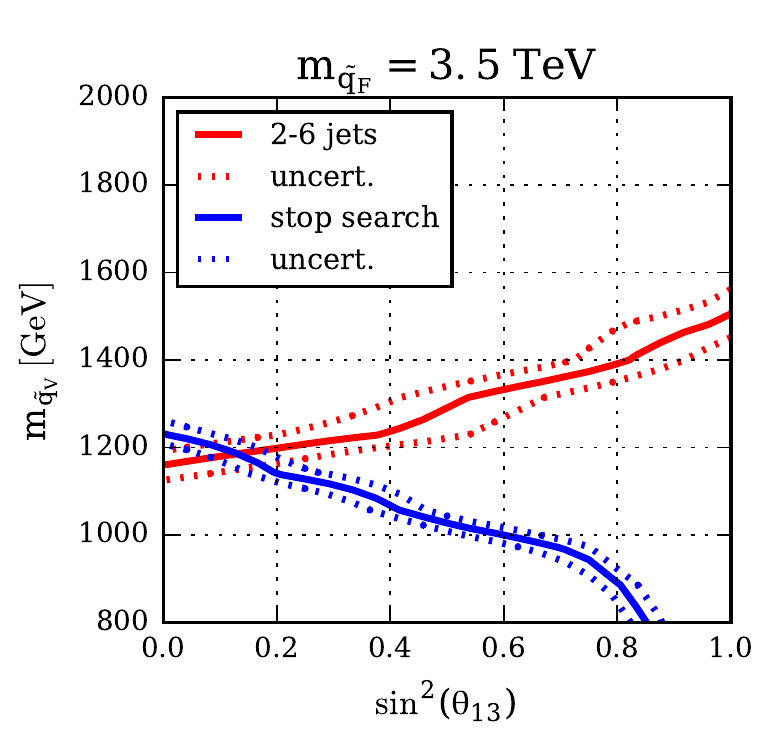}}
\subfloat[\label{fig:4b}]{\includegraphics[width=0.49\textwidth]{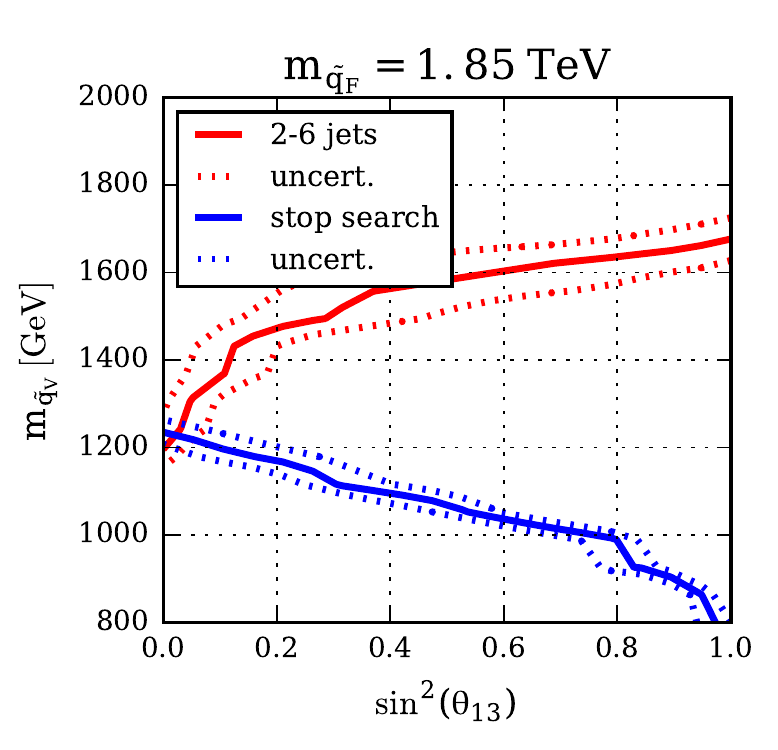}}
\caption{
Mass limits on squarks in case of flavour mixing between 1st and 3rd generation (assuming left-right universality).
For details on the definition of $q_{\tilde V}$ and $q_{\tilde F}$ see eq.~\eqref{eq:states_definiton}.
}
\label{fig:scenarioC}
\end{figure}
In the MRSSM, flavour mixing between the squarks is particularly well motivated \cite{Kribs:2007ac}.
Flavour mixing can modify the limits on squark masses since LHC searches have a different sensitivity to squarks of the first or second generation and to stops and sbottoms.
In the MSSM, the limits on stop masses alone from the dedicated ATLAS stop search~\cite{ATLAS-CONF-2017-019,Aaboud:2017ayj} amount to $m_{\tilde{t}}>1.1$ TeV.
We found that generic squark searches would lead to a limit which is almost equally strong while limits on stops are essentially equal in the MSSM and the MRSSM.

In the following we investigate the impact of flavour mixing on squark mass limits in the MRSSM.
We focus on a simple case exhibiting a particularly strong effect.
We assume equal mixing in the left- and right-handed sectors between first- and third-generation squarks.
As before the LSP remains massless and the gluino mass is $m_{\tilde{g}}=5$ TeV.
The varying squark mass state $q_V$ and fixed squark mass state $q_F$ are related to the ones of the first and third generation via a mixing angle $\theta_{13}$
\begin{align}
\label{eq:states_definiton}
  \tilde{q}_V (\text{lighter
    squark type}) &=\cos\theta_{13}\tilde{q}_3+\sin\theta_{13}\tilde{q}_1\notag\\
  \tilde{q}_F (\text{heavier
    squark type})&=-\sin\theta_{13}\tilde{q}_3+\cos\theta_{13}\tilde{q}_1
\end{align}
With this assignment $\theta_{13}=0$  ($\theta_{13}=\pi/2$ ) corresponds to $\tilde{q}_V\equiv$~stop/sbottom ($\tilde{q}_V\equiv$~sup/sdown).
Figure~\ref{fig:scenarioC} shows the exclusion limit dependence on the mixing angle.

Figure\ \ref{fig:4a} shows the limits on the lighter squark mass $m_{\tilde{q}_V}$ for the case $m_{\tilde{q}_F}=3.5$ TeV, which is an interesting value between the light squark and the gluino masses.
In this case the limit on the lighter squark clearly interpolates between the pure third generation and the pure first generation cases.
For $\theta_{13} = 0$ squark $\tilde{q}_V$ is purely stop-like, and accordingly both kind of limits are around 1.2~TeV.
For higher $\theta_{13}$ the stop content goes down and the stop-search limit becomes weaker while the 2-6 jet limit goes up to $1.5$~TeV.

Figure\ \ref{fig:4b} shows the limits on the lighter squark mass for the case $m_{\tilde{q}_F}=1.85$ TeV, just slightly heavier than the lighter squark.
In this case the heavier squark already almost saturates the limit from generic squark searches discussed above.
For this reason, the lighter squark production easily violates the bounds from generic squark searches limit, and the limit on the lighter squark mass behaves similarly to the case of figure~\ref{fig:4a} but is slightly stronger and goes up to 1.7~TeV.

\section{Conclusions and outlook}

\begin{figure}
\centering
\includegraphics[width=0.49\textwidth]{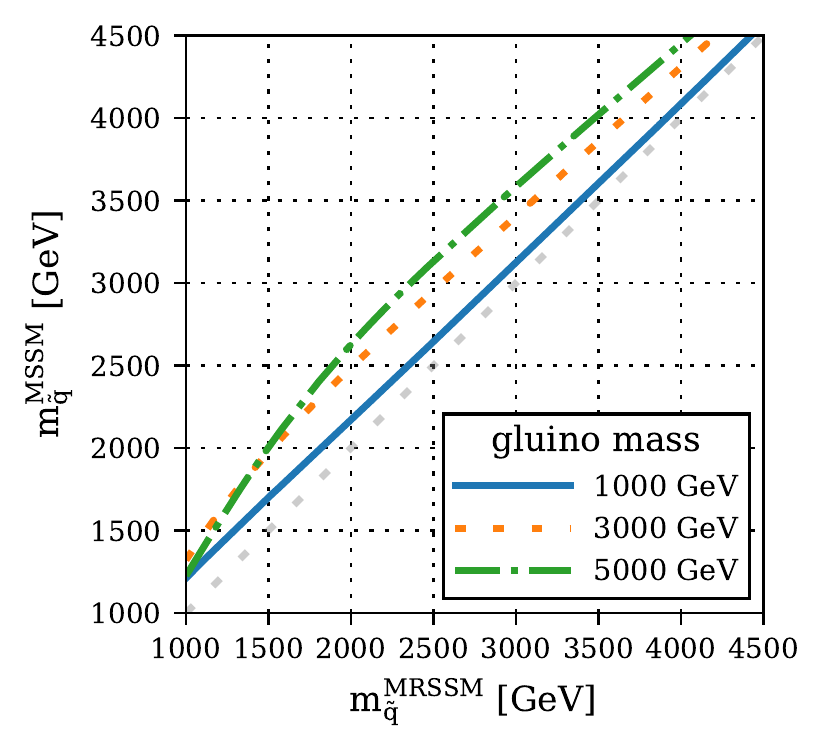}
\caption{
Contours in the MSSM squark mass vs. the MRSSM squark mass plane for masses for which cross sections for 1st and 2nd generation squark production are equal in both models.
}
\label{fig:xsec_comparison}
\end{figure}
R-symmetry is the basis for an exciting alternative to the Minimal Supersymmetric Standard Model, motivated by the suppression of unwanted flavour-violating processes and natural explanation of the non-observation of gluinos.
As it turns out, its other appealing feature is the weakening of the exclusion limits on the pair production of squarks.
In this work we considered 3 distinct, phenomenologically well motivated scenarios.
The case of light first and second generation squarks (with and without mass degeneracy between ``left''- and ``right''-handed squarks) and the case of mixing between stops and up squarks.
Our first main result is that the squark mass limits coming from LHC searches are universally weaker than corresponding MSSM ones.
Specifically we find an 500 -- 600~GeV lower mass limit on squarks if gluinos are heavy (see eqs.~\eqref{eq:first_second_gen} and \eqref{eq:first_second_gen2}).
The second result of the present paper is that flavour mixing between 1st and 3rd generation does not lead to a further dramatic reduction of limits.
The reason is that independently of flavour mixing angle the combination of generic squark searches and stop searches leads to similar limits.

These large differences between the mass limits found in the MRSSM and the MSSM are mostly due to the basic fact that the squark production cross section is significantly lower in the MRSSM.
This is seen in figure~\ref{fig:xsec_comparison} where we plot squark masses in both models for which the squark production cross sections in both models are the same.
For heavy gluino this confirms the 500 -- 600~GeV difference in squark mass limit seen in figure~\ref{fig:scenarioA}.
Subtleties such as model-dependent analysis efficiencies and acceptances do not play a significant role.
Therefore our results are rather robust.
It can be expected that even with future LHC data, if there is no discovery, the mass limits will go up in both models, but the gap between the limits will remain.
For example, the high-luminosity phase of the LHC with an integrated luminosity of 3000 fb$^{-1}$ is expected to be able to exclude MSSM squark masses up to $\sim$3.5 TeV~\cite{ATL-PHYS-PUB-2014-010}; according to our results this translates into an expected limit of only $\sim$3 TeV in the MRSSM (assuming gluino has a mass of 4.5 TeV).
On the other hand the pure gluino mass limits are stronger in the MRSSM as there are twice as many gluino degrees of freedom in the MRSSM.

In deriving those conclusions, we have applied available higher order SQCD corrections to squark pair production and for the remaining processes where such correction are not available, estimated them using MSSM results.
We also estimated the remaining theoretical uncertainty, showing that our conclusion should remain valid even once full, MRSSM specific, NLO+NNLL corrections become available.

{\em Acknowledgements:} This research was supported in part by the German Research Foundation (DFG) under grants number STO 876/4-1 and STO 876/2-2 and the Polish National Science Centre Harmonia grant under contract UMO-2015/18/M/ST2/00518 (2016-2019).

\bibliographystyle{JHEP}
\bibliography{bibliography}

\end{document}